\documentclass[12pt,letterpaper]{article}
\usepackage[letterpaper,margin=1in]{geometry}
\usepackage{microtype}
\usepackage{amssymb,amsmath}
\usepackage{graphicx}
\usepackage{booktabs}
\usepackage{tabularx}
\usepackage{array}
\usepackage{psfrag}
\usepackage{xcolor}
\usepackage{authblk}
\usepackage[numbers,super,sort&compress]{natbib}
\usepackage{setspace}
\usepackage{lineno}
\usepackage{lipsum}

\begin{document}

\title{Exploring Energy Landscapes for Functional Materials: Ga vacancies in $\beta$-Ga$_{2}$O$_{3}$}

\author[1]{Soon Woo Park}
\author[2]{Philipp Pracht}
\author[1]{Se\'an R. Kavanagh\thanks{Corresponding author. Email: sk2045@cam.ac.uk}}
\author[1]{David J. Wales\thanks{Corresponding author. Email: dw34@cam.ac.uk}}
\affil[1]{Yusuf Hamied Department of Chemistry, University of Cambridge, Lensfield Road, Cambridge CB2 1EW, United Kingdom}
\affil[2]{Interdisciplinary Center for Scientific Computing, Ruprecht-Karls-University Heidelberg, Im Neuenheimer Feld 205, 69120 Heidelberg, Germany}
\date{}
\maketitle

\onehalfspacing
\begin{abstract}
Energy landscape exploration requires extensive energy and force evaluations, making efficient access to interatomic potentials an important practical challenge. We present a persistent socket interface that connects the landscape exploration tools in the GMIN, OPTIM, and PATHSAMPLE programs to potentials available through the Atomic Simulation Environment (ASE). The persistent design keeps both the calculator and socket connection active throughout the search, avoiding repeated initialisation. We apply this framework to understand and predict the properties of Ga vacancies in $\beta$-Ga$_2$O$_3$ using two NequIP-based potentials, and construct kinetic transition networks describing complex vacancy configurations and migration pathways. Temperature-dependent analysis of the dominant paths reveals how the most probable migration sequence can change, with the mechanism governed by global properties of the network, rather than simply the highest energy barrier. Visualisation, mean first-passage times, and heat capacities enable us to connect the organisation of the landscape to relaxation time scales and thermodynamic observables. The interface extends energy landscape methods to ASE-compatible atomistic models, providing a general framework for combining global optimisation, transition state searches, kinetic network analysis, and thermodynamic characterisation with recent developments in machine-learned interatomic potentials.
\end{abstract}

\doublespacing

\section{Introduction}
Structural transformations in atomic and molecular systems often proceed through a sequence of metastable configurations. The energy landscape framework\cite{Wales03,Wales18,Schon24} describes these processes in terms of local minima on the potential energy surface (PES) and the transition states that connect them. A collection of these stationary points and their connections provides a network representation of the accessible configurations and the pathways between them. Assigning rate constants to these connections yields a kinetic transition network (KTN), which can be used to analyse competing pathways,  relaxation timescales, and thermodynamic observables.\cite{Wales02}

Constructing this network requires repeated searches for minima and transition states, making the cost of energy and force evaluations a practical constraint on landscape exploration. Data-driven (machine-learned) interatomic potentials (MLIPs) offer an efficient way to approximate quantum mechanical energies and forces, extending the range of systems accessible to atomistic simulation.\cite{BehlerP07,BatznerMSG22,TanDKM26} Many of these models are available through the Atomic Simulation Environment (ASE),\cite{HjorthLarsenJBC17} a Python framework for constructing, manipulating, and simulating atomic systems. Within ASE, energy and force evaluations are handled through a common calculator interface that supports electronic structure codes and interatomic potentials alike. Coupling landscape search algorithms to this interface therefore allows a wide variety of atomistic models to be used within the same exploration framework.

The Cambridge Energy Landscapes software provides a wide range of tools for constructing and analysing energy landscapes. GMIN\cite{GMIN} searches for low-energy structures using basin-hopping global optimisation,\cite{lis87,WalesD97,WalesS99} OPTIM\cite{OPTIM} locates transition states and connecting pathways, and PATHSAMPLE\cite{PATHSAMPLE} coordinates discrete path sampling\cite{Wales02} and kinetic analysis of the resulting KTN.\cite{SharpeW21c,WoodsW24} In the present contribution we describe a persistent socket interface that couples GMIN and OPTIM to ASE calculators running in separate Python processes. The calculator and its connection to the search algorithm are retained across successive evaluations. This interface makes ASE-compatible potentials available for global optimisation and transition state searches, including the OPTIM calculations used by PATHSAMPLE to expand a stationary point database.
In particular, we exploit the tools available in OPTIM to characterise multistep paths via the missing connection algorithm\cite{CarrTW05}
to optimise complex defect migration pathways within a growing KTN, with an initial exploration based on
various basin-hopping approaches in GMIN.

We apply the interface to analyse Ga vacancies in $\beta$-Ga$_{2}$O$_{3}$, an ultrawide-band-gap semiconductor of interest for power electronics. Intrinsic defect migration in this material contributes to atomic transport and can mediate the redistribution of dopants and impurities.\cite{FrodasonVJV23} Ga vacancies are of particular interest because they act as compensating acceptors and can undergo substantial atomic reconstruction, which affects their stability and electrical behaviour.\cite{varley_hydrogenated_2011,Kavanagh25,mosquera-lois_search_2021} An important example of such reconstruction is the split vacancy, in which a neighbouring Ga atom moves away from its lattice site towards the original vacancy, producing a configuration with two vacant lattice sites and an intervening Ga atom.\cite{varley_hydrogenated_2011,Kavanagh25} Such configurations have been identified experimentally by positron annihilation spectroscopy,\cite{KarjalainenMEG21} with additional evidence from hydrogen-related vibrational spectroscopy.\cite{Weiser18} This reconstruction significantly lowers the energy of the cation vacancy defect in $\beta$-Ga$_{2}$O$_{3}$, limiting $n$-type conductivity and altering migration pathways. 
Recent searches using electrostatic screening and MLIPs have demonstrated the presence of such `split vacancy' configurations across a broad range of materials.\cite{Kavanagh25} 
Beyond split vacancies, many other defect reconstruction motifs have been observed,\cite{mosquera-lois_identifying_2023} further highlighting the need for energy landscape exploration to consider complex defect geometries.

Vacancy reconstruction also has direct consequences for migration. For example, the $V_{\mathrm{Ga}}^{3-}$ charge state can migrate between split configurations via `three-split' intermediates, which lower the migration barrier along the [001] direction.\cite{FrodasonVJV23} Diffusion networks have also been constructed from split-defect configurations and their connecting hops, with master diffusion equations used to obtain directional diffusivities.\cite{LeeSE24} These studies establish the importance of multiple defect configurations and connected migration pathways. Here, we focus on how branching at intermediate minima changes the relative probabilities of competing migration pathways with temperature. For a multistep route, the probability of following a particular sequence depends on the relative rates of all the possible exits from each intermediate minimum. Analysing these branching probabilities alongside the potential energy barriers allows changes in mechanism to be related to the surrounding network.

In this work, we use the GMIN/OPTIM--ASE interface to construct and analyse a Ga-vacancy landscape described by pretrained NequIP-based ML potentials.\cite{KimWVZ25,KavanaghTWD26} Comparisons with density functional theory are used to assess the predicted relative stability of representative defect structures. We classify migration routes in terms of recurring structural mechanisms and compare their energy profiles and branching probabilities to determine how routes selected by Dijkstra analysis change with temperature. We then use disconnectivity graphs to examine the organisation of point, split, and three-split vacancies across the landscape, and mean first-passage times (MFPTs) to quantify relaxation to low-energy states. Equilibrium populations and heat capacities, evaluated within the harmonic superposition approximation (HSA),\cite{Mcginty71, Burton72, Hoare79, StillingerW84, FrankeHB93,Wales03} characterise the redistribution among structural families with temperature. This application demonstrates how the socket interface extends established energy landscape tools to ASE-compatible models, while retaining structural, kinetic, and thermodynamic analyses within a common stationary point framework.

\section{Methods}
\subsection{Socket-based coupling of the Cambridge Energy Landscapes software and ASE}

\begin{figure*}[!t]
  \centering
  \includegraphics[width=0.98\textwidth]{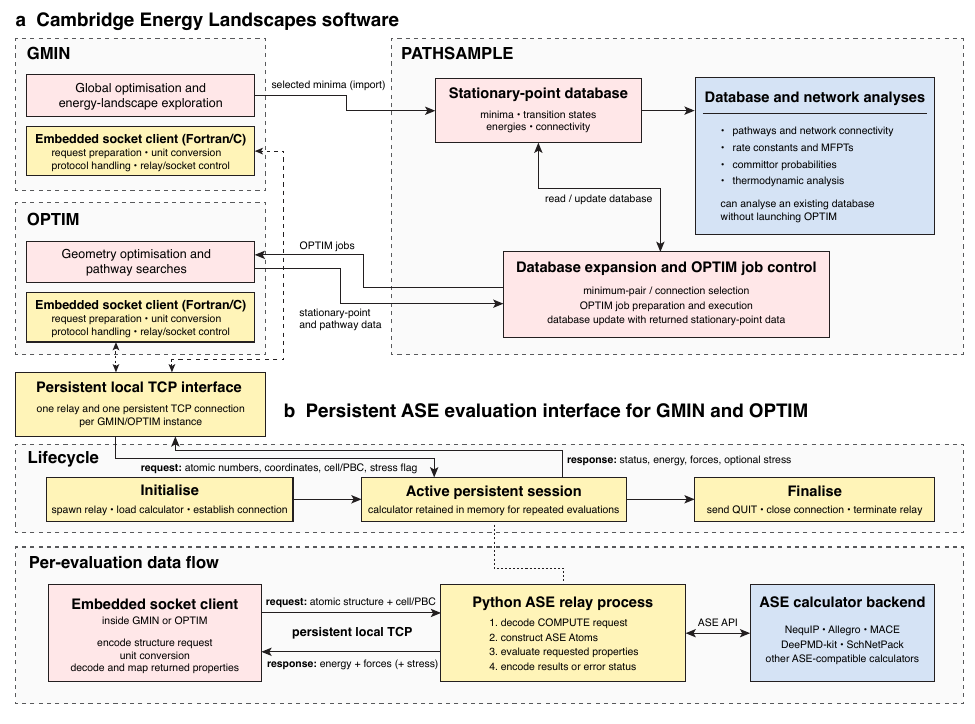}
  \caption{Workflow coupling the Cambridge Energy Landscapes software to ASE calculators through a persistent socket interface. (a) GMIN supplies low-energy minima, while PATHSAMPLE coordinates OPTIM searches to construct and expand a stationary-point database for pathway, kinetic, and thermodynamic analyses. (b) Persistent ASE evaluation interface used by each GMIN or OPTIM process. An embedded Fortran/C client communicates with a Python relay over a local TCP connection, which calls the selected ASE calculator and returns energies, forces, and, when required, stresses. The calculator and connection are retained for repeated evaluations until the process terminates.}
  \label{fig:workflow}
\end{figure*}

We developed a persistent socket interface coupling GMIN/OPTIM to ASE\cite{HjorthLarsenJBC17} calculators by adapting and extending Pracht's open-source \texttt{fmlip-relay} implementation\cite{Pracht26} (Fig.~\ref{fig:workflow}). Each GMIN/OPTIM process launches a separate Python relay that loads the selected calculator and communicates with an embedded Fortran/C client over a local TCP connection. Both the calculator and connection are maintained for the lifetime of the GMIN/OPTIM process, avoiding repeated model initialisation between evaluations. This persistent architecture allows the landscape search algorithms and ASE-based calculators to operate in their native software environments.

For each evaluation, the interface passes the atomic species, Cartesian coordinates, cell vectors and periodic boundary conditions to the calculator and returns the potential energy and atomic forces. For variable-cell calculations, the calculator also supplies the stress tensor. In this case, the fractional coordinates used internally by GMIN and OPTIM are converted to Cartesian coordinates before evaluation, and the returned forces and stress tensor are transformed into the atomic and cell derivatives required by the optimisers. Calculator-specific operations are encapsulated in a backend layer, separate from socket communication and process management. A shared evaluation routine obtains energies, forces and optional stress tensors through ASE, while each backend specifies how its calculator is loaded and configured. Additional ASE-compatible calculators can therefore be incorporated through this backend layer without modifying the socket protocol or the landscape search algorithms.

The interface can be used for independent energy-landscape exploration or for database construction coordinated by PATHSAMPLE. For example, selected minima from GMIN can be imported into a PATHSAMPLE database and connected through OPTIM pathway searches. During database expansion, PATHSAMPLE launches independent OPTIM jobs, each using its own persistent connection to an ASE relay, and incorporates the returned stationary points and their connectivity into the database. Once the database has been constructed, kinetic and thermodynamic analyses can be performed directly from the stored stationary-point data without further potential evaluations.

A bytewise comparison confirmed that the energy, force, and stress data transmitted by the Python relay were received unchanged by the Fortran client. Timing measurements were collected during a GMIN run for a 99-atom supercell, with the NequIP-OAM-L model evaluated on the GPU through ASE while GMIN ran on the CPU. For single-point energy and force evaluations, the difference between the median GMIN round-trip time and the median Python backend evaluation time was 0.2\,ms, corresponding to 0.4\% of the median backend evaluation time. Complete timing statistics are reported in Table~S1 of the Supporting Information.

\subsection{Ga-vacancy model and reference calculations}

A Ga vacancy was introduced by removing one Ga atom from a 100-atom supercell of $\beta$-Ga$_{2}$O$_{3}$. Initial Ga-vacancy configurations were prepared using the \texttt{doped} package\cite{KavanaghSNM24}, with structural manipulation and symmetry analysis performed using \texttt{pymatgen}\cite{OngRJH13}. All energy landscape calculations were performed for periodic boundary conditions with a fixed cell defined by $a=b=10.61$\,\AA{}, $c=11.63$\,\AA{}, $\alpha=\beta=114.24^{\circ}$, and $\gamma=91.78^{\circ}$, giving a minimum image distance of 10.61\,\AA{}. For these calculations, two NequIP-based potentials were separately tested to evaluate energies and forces: the pretrained NequIP-OAM-L\cite{KavanaghTWD26} and an OMat24-trained NequIP-XL model augmented with latent Ewald summation (LES),\cite{KimWVZ25} denoted NequIP-OMat24-XL-LES (hereafter NequIP-LES). The latter form incorporates LES, in which latent variables learned from energy and force data are used to represent long-range interactions.\cite{Cheng25} No explicit defect charge or atomic partial charges were supplied as inputs. Separate stationary point databases were constructed using the two potentials to assess the sensitivity of the predicted vacancy landscape and migration behaviour to the choice of potential. Unless otherwise stated, the detailed energy landscape, kinetic, and thermodynamic analyses presented below use NequIP-LES; the NequIP-OAM-L migration-route profiles and summary statistics are provided in the Supporting Information.

Reference Density Functional Theory (DFT) calculations were performed using the Vienna \textit{Ab initio} Simulation Package (VASP)\cite{KresseF96} within the projector augmented-wave (PAW) framework.\cite{KresseJ99} Exchange and correlation were described using the PBEsol functional.\cite{PerdewRCV08} Representative point, split, and three-split configurations obtained from the NequIP-OAM-L landscape were relaxed in the $V_{\mathrm{Ga}}^{3-}$ charge state using a plane-wave cutoff energy of 500\,eV and $\Gamma$-point reciprocal space sampling. The atomic coordinates were optimised with the lattice vectors fixed until the residual forces on all atoms were below 0.01\,eV\,\AA{}$^{-1}$.

\subsection{Energy landscape exploration and vacancy classification}

Low-energy minima were located by basin-hopping global optimisation\cite{WalesD97} using GMIN.\cite{GMIN} Each trial move consisted of random Cartesian displacements followed by local minimisation, and acceptance was determined by the Metropolis criterion applied to the energies of the resulting minima. The search comprised 10,000 basin-hopping steps at a fictitious temperature of $k_{\mathrm{B}}T=0.0260$\,eV for the accept/reject condition, with displacements sampled uniformly up to a maximum change of 0.3\,\AA{} in each Cartesian component. Local minimisation was performed using a customised limited-memory Broyden--Fletcher--Goldfarb--Shanno (L-BFGS)\cite{Broyden70,Fletcher70,Goldfarb70,Shanno70,LiuN89} algorithm, with convergence defined as a root-mean-square (RMS) force below $10^{-5}$\,eV\,\AA{}$^{-1}$. To remove duplicate entries, minima differing in energy by less than $10^{-4}$\,eV were treated as equivalent and only one was retained. The 100 lowest-energy retained minima were then used to seed the stationary point database.

The database was expanded by discrete path sampling,\cite{Wales02,Wales04} which constructs a network of local minima linked by transition states. PATHSAMPLE selected pairs of minima for connection attempts and incorporated the stationary points returned by OPTIM. Initial paths between the selected endpoints were investigated using the doubly-nudged\cite{TrygubenkoW04,SheppardTH08} elastic band\cite{Millsjs95,JonssonMJ98,HenkelmanUJ00,HenkelmanJ00} (DNEB) method. 
Only images that are local maxima in the chain are of interest in this phase, so it is only necessary to optimise the band until these points stand out consistently.
These candidates for transition states were then refined by hybrid eigenvector-following,\cite{munrow99,henkelmanj99,kumedamw01,ZengXH14} using a Rayleigh--Ritz procedure to determine the uphill search direction. The resulting transition states and their connected minima were refined to an RMS force below $10^{-5}$\,eV\,\AA{}$^{-1}$. To further expand the database, additional pairs of minima were selected for connection using the Dijkstra-based\cite{dijkstra59} missing-connection algorithm.\cite{CarrTW05} Two stationary points were treated as identical when both their energy difference and total atomic displacement after structural alignment were below $10^{-4}$\,eV and 0.4\,\AA{}, respectively. The latter parameter is the Euclidean norm of all atomic displacements, minimised over the allowed atom permutations using periodic minimum-image distances. Database expansion focused on connecting the low-energy region and the 34 endpoint pairs used in the subsequent analyses.

Disconnectivity graphs\cite{beckerk97,walesmw98} were used to visualise the barrier structure of the sampled landscape. Minima connected below a given energy threshold are grouped into the same superbasin, and as the threshold energy increases in regular steps, the merging of superbasins reveals the organisation of the landscape into funnels and the barriers between them. An energy spacing of 0.1\,eV was used for the threshold in superbasin analysis for both the full and magnified disconnectivity graphs.
We note that other representations of the landscape could also be used.\cite{SchonPJ96,Heuer97}

Vacancy configurations were classified by comparing the lattice Ga-site occupancies of each minimum with the pristine structure. After aligning the structures using the O sublattice under periodic boundary conditions, the \texttt{doped} package\cite{KavanaghSNM24} was used to identify unoccupied Ga lattice sites and additional (interstitial) Ga atoms. The site-matching tolerance was set to 0.919\,\AA{}, corresponding to half the shortest bond length in the bulk structure (1.838\,\AA{}). One unoccupied Ga site with no off-site Ga atom defines a point vacancy; two unoccupied sites with one interstitial Ga atom define a split vacancy; and three unoccupied sites with two interstitial Ga atoms define a three-split vacancy. Alternative occupancy patterns were grouped as ``other''. Four-split configurations, characterised by four unoccupied Ga lattice sites and three off-site Ga atoms, were counted separately in the database-wide classification reported in Table~S2 of the Supporting Information, but retained within the ``other'' category in the thermodynamic analysis.

\subsection{Kinetic network and pathway analysis}

Migration pathways were analysed for all 34 symmetry-inequivalent endpoint pairs involving vacancies on octahedral and tetrahedral Ga sites. Dijkstra-based searches were performed at $k_{\mathrm{B}}T=0.0260$, 0.0345, and 0.0689\,eV, corresponding approximately to 300, 400, and 800\,K.
Here, the edge weight is set to (minus) the natural logarithm of the branching probability, so that the shortest paths correspond to the largest contributions in the rate constant expression when intervening minima are treated in a steady-state approximation.\cite{Wales02}

Approximate transition rates between connected minima were evaluated using harmonic transition state theory in the canonical ensemble.\cite{Wales06} Hessians were calculated numerically in OPTIM by finite differences of the gradients with Richardson extrapolation, using a displacement parameter of $10^{-4}$\,\AA{}. The Hessians were then mass-weighted and diagonalised to obtain the normal-mode angular frequencies. The elementary rate constant for a transition from minimum $\alpha$ to minimum $\beta$ through a connecting transition state is
\begin{equation}
k_{\beta\alpha}(T)
=
\sigma
\frac{(\bar{\nu}_{\alpha})^{\kappa}}
     {(\bar{\nu}_{\beta\alpha})^{\kappa-1}}
\exp\left[
-\frac{V_{\beta\alpha}-V_{\alpha}}{k_{\mathrm B}T}
\right].
\end{equation}
Here, $V_{\alpha}$ and $V_{\beta\alpha}$ are the energies of the minimum and transition state, respectively; $\bar{\nu}$ denotes the geometric mean of the positive vibrational frequencies ($\nu=\omega/2\pi$), $\kappa$ is the number of vibrational degrees of freedom at the minimum, and $\sigma$ is the symmetry factor. The three translational zero frequency modes and the unstable mode at each transition state were excluded from these products, as usual.

For a transition from minimum $j$ to minimum $i$, the total rate $k_{ij}(T)$ was obtained by summing the rates through all transition states connecting the pair. The branching probability that the next transition from $j$ reaches $i$ is then
\begin{equation}
B_{ij}(T)=\frac{k_{ij}(T)}{\displaystyle\sum_{\ell\ne j}k_{\ell j}(T)},
\end{equation}
where the denominator includes all exits from $j$ to other minima in the network. Assigning an edge cost of $-\ln B_{ij}(T)$ allows Dijkstra's algorithm\cite{dijkstra59} to identify the route $\Gamma$ with the smallest path action, $\mathcal{S}_{\Gamma}(T)$,\cite{Wales02,SharpeW21a}
\begin{equation}
\label{eq:weight}
\mathcal{S}_{\Gamma}(T)
=-\ln \mathcal{W}[\Gamma;T]
=-\sum_{(j\rightarrow i)\in\Gamma}\ln B_{ij}(T),
\qquad
\mathcal{W}[\Gamma;T]=\prod_{(j\rightarrow i)\in\Gamma}B_{ij}(T).
\end{equation}
Here, $\mathcal{W}[\Gamma;T]$ is the path weight, corresponding to the probability of following the specified sequence of transitions given its initial minimum. The selected route thus maximises this probability;\cite{Wales02,SharpeW21a} it need not have the fewest steps, the lowest maximum energy barrier, or the shortest mean passage time.

Inspection of the migration routes identified seven recurring migration motifs, denoted $\mathrm{M}_{1}$--$\mathrm{M}_{7}$. A motif can contain more than one minimum--transition state--minimum segment, and a superscript $-1$ (e.g., $\mathrm{M}_i^{-1}$) denotes traversal in the reverse direction. 

Three endpoint pairs (paths 5, 14, and 29) exhibiting distinct temperature-dependent route changes were selected for a detailed comparison of competing migration routes. For these pairs, the path actions of the distinct routes selected at approximately 300, 400, and 800\,K were evaluated from 300 to 800\,K in 5\,K increments. Crossings of the resulting path-action curves identify changes in the preferred route among these candidates. To illustrate these changes, the routes at representative temperatures of 300, 400, and 800\,K were further compared using their potential energy profiles as a function of integrated path length for the approximate steepest-descent path. The integrated path length, $s$, was calculated by accumulating Euclidean distances between successive configurations along the discretised approximate steepest-descent pathways, accounting for periodic boundary conditions.

Relaxation to the lowest-energy region was quantified by mean first-passage times (MFPTs), using graph transformation with a backward pass in PATHSAMPLE.\cite{Wales09,SharpeW21a,SharpeW21b} This procedure eliminates intermediate states via renormalisation of branching probabilities and waiting times, preserving the MFPTs to the absorbing set, then recovers the MFPTs for the starting minima during the backward pass. The 48 minima in the lowest-energy region were treated as the absorbing set $A$. For each starting minimum, the MFPT is the expected time to reach any member of $A$ for the first time. Calculations used the same harmonic rate model and the three temperatures specified above.

\subsection{Equilibrium thermodynamics}

Equilibrium populations and heat capacities were calculated within the classical HSA.\cite{Mcginty71, Burton72, Hoare79, StillingerW84, FrankeHB93,Wales03} In this approximation, each minimum contributes a harmonic basin to the partition function, with a contribution proportional to
\begin{equation}
Z_i(T)\propto
\frac{1}{o_i}
\exp\left(-\frac{E_i}{k_{\mathrm{B}}T}\right)
\prod_{\alpha=1}^{\kappa}\frac{1}{\omega_{i\alpha}},
\end{equation}
where $E_{i}$ is the potential energy, $o_{i}$ is the point-group order, and $\omega_{i\alpha}$ are the non-zero normal-mode frequencies of minimum $i$. Factors common to all minima were omitted because they cancel in the occupation probabilities. The occupation probability $p_{i}(T)$ and the population $P_{c}(T)$ of structural class $c$ are then
\begin{equation}
p_{i}(T)=\frac{Z_{i}(T)}{\displaystyle\sum_{j} Z_{j}(T)},
\qquad
P_{c}(T)=\sum_{i\in c}p_{i}(T).
\end{equation}

The constant volume heat capacity separates into a common harmonic term and a contribution from redistribution among minima:
\begin{equation}
\frac{C_{V}(T)}{k_{\mathrm{B}}}
=\kappa+
\frac{\langle\Delta E^2\rangle-\langle\Delta E\rangle^2}
{(k_{\mathrm{B}}T)^2},
\end{equation}
where $\Delta E_{i}=E_{i}-E_{0}$, $E_{0}$ is the global minimum energy, and $\langle X\rangle=\sum_{i} p_{i} X_{i}$. For the fixed-cell periodic system, removal of the three translational modes leaves $\kappa=3N-3$ vibrational degrees of freedom. To identify the structural families contributing to population redistribution, the second term was decomposed by structural class, defining the contribution of class $c$ as
\begin{equation}
\frac{C_{\mathrm{land},c}(T)}{k_{\mathrm{B}}}
=\sum_{i\in c}p_i(T)
\left[\frac{\Delta E_i-\langle\Delta E\rangle}{k_{\mathrm{B}}T}\right]^2.
\end{equation}
These non-negative contributions sum to $C_V/k_{\mathrm{B}}-\kappa$ when all classes are included. Because they use the mean energy of the full ensemble, they measure each class's contribution to the total heat capacity rather than the heat capacity of that class in isolation.

The minima involved in thermal redistribution were identified from the temperature derivative of their occupation probabilities,\cite{Wales17}
\begin{equation}
\frac{\partial p_i}{\partial T}
=\frac{p_i(T)}{k_{\mathrm{B}}T^2}
\left(\Delta E_i-\langle\Delta E\rangle\right).
\end{equation}
The sign of $\partial p_i/\partial T$ distinguishes minima that gain population as $T$ changes from those that lose population. At each temperature examined, minima within each sign group were ranked by their individual contributions to the landscape-dependent heat capacity. The leading contributors accounting cumulatively for 99\% of each group's contribution were selected for display. This selection was used only to highlight relevant minima; equilibrium populations and heat capacities were always evaluated over the full database.

\section{Results and discussion}
\subsection{Vacancy configurations and relative stability}

\begin{figure}[!t]
  \centering
  \includegraphics[width=0.98\columnwidth]{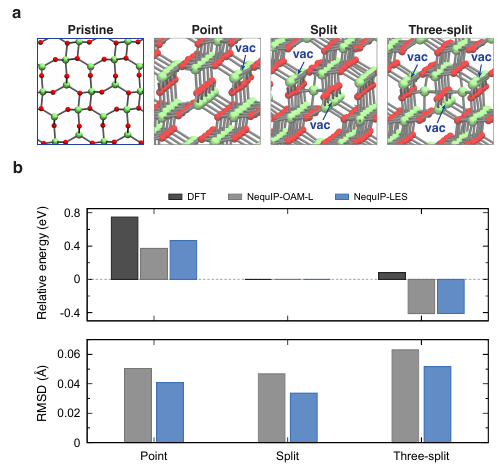}
  \caption{Representative Ga-vacancy configurations and their relative stability. (a) Pristine $\beta$-Ga$_{2}$O$_{3}$ and the lowest-energy point, split, and three-split Ga-vacancy configurations. Green and red spheres denote Ga and O atoms, respectively, and blue labels indicate vacant Ga lattice sites. (b) Relative energies obtained using DFT, NequIP-OAM-L, and NequIP-LES (top), and the root-mean-square deviations (RMSDs) of the corresponding MLIP structures from the DFT-relaxed structures (bottom). For each method, the relative energies are referenced to the split-vacancy configuration. The DFT-relaxed structures serve as the RMSD reference and are therefore not included in the lower graph.}
  \label{fig:structures}
\end{figure}

A Ga vacancy formed by removing a single Ga atom from $\beta$-Ga$_{2}$O$_{3}$ can adopt several configurations [Fig.~\ref{fig:structures}(a)], including point, split, and three-split vacancies.\cite{FrodasonVJV23} These structures all have the same total composition: they are distinguished by how the surrounding atoms rearrange to accommodate the missing Ga atom. In a point vacancy, the unoccupied site remains readily identifiable, whereas reconstruction distributes the defect over several lattice sites. The structural families therefore describe different arrangements of a single Ga vacancy, rather than different numbers of removed atoms.

Reconstruction substantially changes the relative stability of the vacancy configurations [Fig.~\ref{fig:structures}(b)]. Using the split vacancy as the energy reference for each method, DFT places the lowest-energy point and three-split configurations 0.75 and 0.08\,eV higher in energy, respectively. Both MLIPs predict lower relative energies for these configurations than DFT. The corresponding relative energies are 0.37 and $-0.41$\,eV with NequIP-OAM-L, and 0.47 and $-0.41$\,eV with NequIP-LES. Hence, although the point vacancy remains less stable than the split vacancy, the three-split configuration becomes the global minimum in both MLIP landscapes. One possible contribution to this reversal is the different treatment of the charged defect. The DFT calculations explicitly describe $V_{\mathrm{Ga}}$ in the $-3$ charge state, whereas neither of the MLIP calculations in this work was fitted to data with an explicit defect charge. Although LES enables NequIP-LES to represent fitted long-range electrostatic interactions, it does not impose the specific charge state of the defective cell. The slightly altered energetic ordering between NequIP-OAM-L and NequIP-LES therefore cannot be attributed uniquely to charge effects, and may also reflect differences in fitting data coverage and uncertainty in the pretrained potentials for defect configurations that were not included in their training data. 
Despite the small energetic discrepancy, both MLIPs reproduce the DFT-relaxed local geometries closely. The root-mean-square deviations (RMSDs) in atomic coordinates are 0.050, 0.047, and 0.063\,\AA{} for NequIP-OAM-L and 0.041, 0.034, and 0.052\,\AA{} for NequIP-LES, for the point, split, and three-split configurations, respectively. NequIP-LES is therefore consistently closer to the DFT-relaxed structures. These results show that both MLIPs retain the characteristic vacancy reconstructions found in the DFT-relaxed structures, despite differences in their relative energies. For the subsequent energy landscape analyses, we focus on NequIP-LES, which gives consistently smaller RMSDs and incorporates long-range interactions. The corresponding NequIP-OAM-L results are provided in the Supporting Information for comparison.

\subsection{Vacancy migration mechanisms and pathways}

\begin{figure*}[!t]
  \centering
  \includegraphics[width=0.98\textwidth]{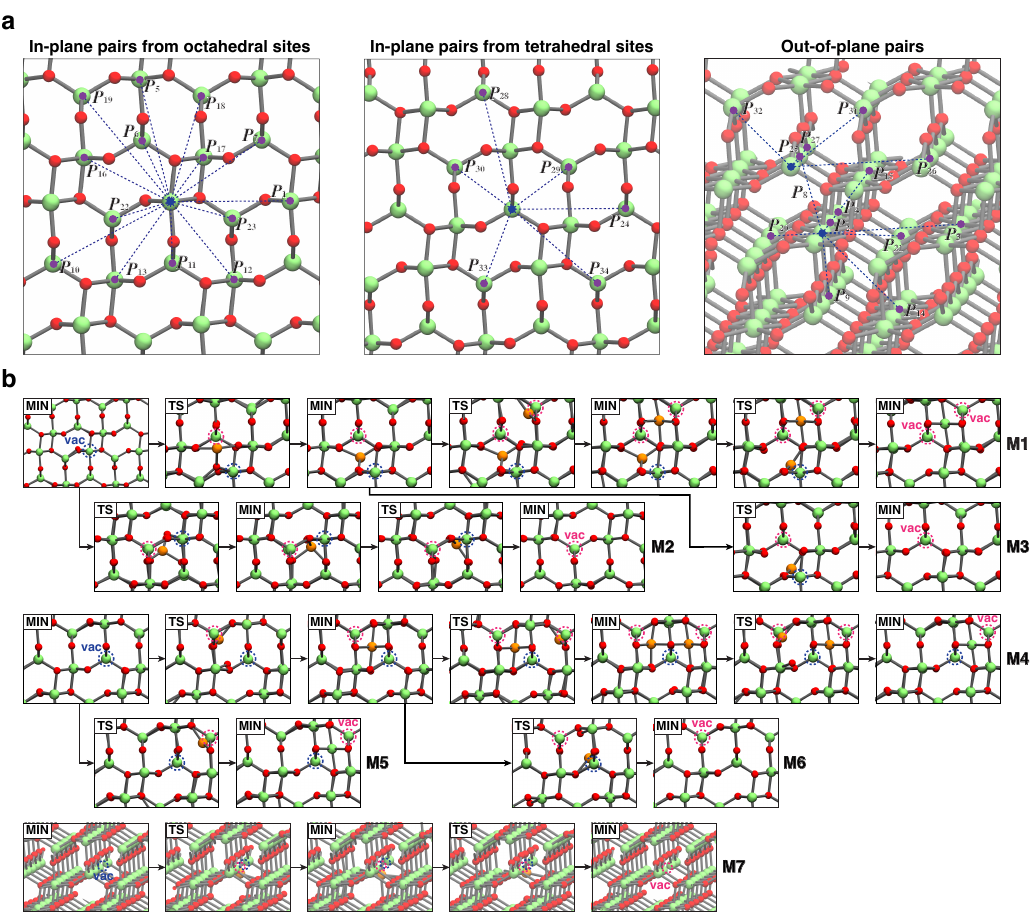}
  \caption{Vacancy endpoint pairs and recurring migration motifs. (a) The 34 pairs numbered 1 to 34, grouped into in-plane connections from octahedral and tetrahedral Ga sites and out-of-plane connections. Dashed lines identify endpoint pairs rather than atomic trajectories. (b) Representative sequences of minima (MIN) and transition states (TS) defining $\mathrm{M}_{1}$--$\mathrm{M}_{7}$. Green and red spheres denote Ga and O atoms, respectively, and orange spheres highlight displaced Ga atoms. Blue and red vacancy-site markers indicate the initial and final vacancy positions, respectively. Shared portions of the sequences are connected by arrows.}
  \label{fig:pathways}
\end{figure*}

The endpoint pairs in Fig.~\ref{fig:pathways}(a) describe migration between point Ga vacancies at different lattice sites. All 780 distinct pairs formed from the 40 Ga sites in the pristine supercell were enumerated and grouped according to the space-group symmetry of the pristine structure, yielding 34 symmetry-inequivalent endpoint pairs. These pairs sample vacancy displacements within and between the Ga-containing planes, starting from both octahedral and tetrahedral sites. Resolving these connections into their constituent stationary points reveals a much richer set of rearrangements than the endpoint geometries alone suggest. All selected pathways can nevertheless be described using the seven recurring motifs in Fig.~\ref{fig:pathways}(b). Some motifs contain several transition states and intervening minima, providing a compact description of multistep structural rearrangements.

For an octahedral-site vacancy, $\mathrm{M}_{2}$ and $\mathrm{M}_{3}$ provide two-step connections to tetrahedral-site vacancies through different intermediate configurations. However, the maximum energy barriers differ significantly: 1.39\,eV for $\mathrm{M}_{2}$ and 0.54\,eV for $\mathrm{M}_{3}$. Consistent with its lower maximum barrier, $\mathrm{M}_{3}$ appears repeatedly in the selected pathways, either at the beginning of a pathway or in reverse at its end. The alternative motif $\mathrm{M}_{1}$ shares the initial rearrangement of $\mathrm{M}_{3}$ but continues through further reconstruction to a split configuration. Notably, $\mathrm{M}_{1}$ appears only in pathways selected at the highest sampled temperature, where it replaces lower-temperature sequences that begin with $\mathrm{M}_{3}$ for several endpoint pairs (Table~\ref{tab:pathways}). Thus, a common initial displacement can branch into different subsequent migration sequences, and the preferred continuation can change with temperature.

\begin{table*}[!t]
  \centering
  \caption{Dijkstra-selected pathways for the 34 vacancy migration routes defined in Fig.~\ref{fig:pathways}. Migration motifs $\mathrm{M}_{1}$--$\mathrm{M}_{7}$ are defined in the same figure, and consecutive mechanism labels denote their sequential occurrence along a pathway. $T_{300}$, $T_{400}$, and $T_{800}$ correspond to $k_{\mathrm{B}}T=0.0260$, 0.0345, and 0.0689\,eV (300, 400, and 800\,K), respectively. The pathways are grouped according to whether the same Dijkstra route is obtained at all three sampled temperatures or the selected route varies between them. Bold $T_i$ labels identify the sampled temperatures when different routes are selected. $\Delta E_{\max}^{\ddagger}$ denotes the largest local forward barrier along the selected pathway, defined as the maximum energy difference between a transition state and its immediately preceding minimum in the direction of traversal. A single value indicates that $\Delta E_{\max}^{\ddagger}$ is the same at all three sampled temperatures to the reported precision; where three values are given, they correspond to $T_{300}/T_{400}/T_{800}$.}
\vspace{6pt}
\label{tab:pathways}

\small
\setlength{\tabcolsep}{4pt}
\renewcommand{\arraystretch}{1.05}

\begin{tabularx}{\textwidth}{@{}
c >{\raggedright\arraybackslash}X r
@{\hspace{2.5em}}
c >{\raggedright\arraybackslash}X r
@{}}

\toprule

\multicolumn{6}{@{}l}{
\textit{Pathways unchanged at the selected temperatures}} \\
\addlinespace[2pt]
\textbf{Path}
& \textbf{Dijkstra pathway}
& $\boldsymbol{\Delta E_{\mathrm{max}}^{\ddagger}}$ \textbf{(eV)}
&
\textbf{Path}
& \textbf{Dijkstra pathway}
& $\boldsymbol{\Delta E_{\mathrm{max}}^{\ddagger}}$ \textbf{(eV)}
\\
\midrule

 2 & $\mathrm{M}_{3}\mathrm{M}_{7}\mathrm{M}_{3}^{-1}$ & 0.855
&  4 & $\mathrm{M}_{3}\mathrm{M}_{7}\mathrm{M}_{7}\mathrm{M}_{3}^{-1}$ & 0.855
\\
 6 & $\mathrm{M}_{3}$ & 0.543
&  8 & $\mathrm{M}_{3}\mathrm{M}_{7}$ & 0.855
\\
 10 & $\mathrm{M}_{3}\mathrm{M}_{7}\mathrm{M}_{7}$ & 0.855
&  13 & $\mathrm{M}_{3}\mathrm{M}_{6}\mathrm{M}_{3}^{-1}$ & 1.093
\\
 15 & $\mathrm{M}_{3}\mathrm{M}_{6}\mathrm{M}_{7}\mathrm{M}_{3}^{-1}$ & 1.093
&  16 & $\mathrm{M}_{2}\mathrm{M}_{3}^{-1}$ & 1.392
\\
17 & $\mathrm{M}_{3}\mathrm{M}_{6}\mathrm{M}_{7}\mathrm{M}_{7}\mathrm{M}_{3}^{-1}$ & 1.093
& 19 & $\mathrm{M}_{3}\mathrm{M}_{6}$ & 1.093
\\
21 & $\mathrm{M}_{3}\mathrm{M}_{6}\mathrm{M}_{7}$ & 1.093
& 22 & $\mathrm{M}_{2}$ & 1.392
\\
23 & $\mathrm{M}_{3}\mathrm{M}_{6}\mathrm{M}_{7}\mathrm{M}_{7}$ & 1.093
& 25 & $\mathrm{M}_{7}$ & 0.855
\\
27 & $\mathrm{M}_{7}\mathrm{M}_{7}$ & 0.855
& 30 & $\mathrm{M}_{6}$ & 1.093
\\
32 & $\mathrm{M}_{6}\mathrm{M}_{7}$ & 1.093
& 34 & $\mathrm{M}_{6}\mathrm{M}_{7}\mathrm{M}_{7}$ & 1.093
\\
\midrule

\multicolumn{6}{@{}l}{
\textit{Pathways varying across the selected temperatures}} \\
\addlinespace[2pt]
\textbf{Path}
& \multicolumn{4}{l}{\textbf{Dijkstra pathway}}
& $\boldsymbol{\Delta E_{\mathrm{max}}^{\ddagger}}$ \textbf{(eV)}
\\
\midrule

1 & \multicolumn{4}{l}{
  $\boldsymbol{T_{300}}$: $\mathrm{M}_{3}\mathrm{M}_{4}\mathrm{M}_{4}^{-1}\mathrm{M}_{3}^{-1}$;
  $\boldsymbol{T_{400}}$: $\mathrm{M}_{3}\mathrm{M}_{5}\mathrm{M}_{4}^{-1}\mathrm{M}_{3}^{-1}$;
  $\boldsymbol{T_{800}}$: $\mathrm{M}_{1}\mathrm{M}_{4}^{-1}\mathrm{M}_{3}^{-1}$
} & 1.116
\\
3 & \multicolumn{4}{l}{
  $\boldsymbol{T_{300,400}}$: $\mathrm{M}_{3}\mathrm{M}_{7}\mathrm{M}_{4}\mathrm{M}_{4}^{-1}\mathrm{M}_{3}^{-1}$;
  $\boldsymbol{T_{800}}$: $\mathrm{M}_{1}\mathrm{M}_{4}^{-1}\mathrm{M}_{7}\mathrm{M}_{3}^{-1}$
} & 1.116
\\
5 & \multicolumn{4}{l}{
  $\boldsymbol{T_{300}}$: $\mathrm{M}_{3}\mathrm{M}_{7}\mathrm{M}_{4}\mathrm{M}_{4}^{-1}\mathrm{M}_{7}\mathrm{M}_{3}^{-1}$;
  $\boldsymbol{T_{400,800}}$: $\mathrm{M}_{2}\mathrm{M}_{6}\mathrm{M}_{3}^{-1}$
} & 1.116/1.392/1.392
\\
7 & \multicolumn{4}{l}{
  $\boldsymbol{T_{300}}$: $\mathrm{M}_{3}\mathrm{M}_{4}\mathrm{M}_{4}^{-1}$;
  $\boldsymbol{T_{400}}$: $\mathrm{M}_{3}\mathrm{M}_{5}\mathrm{M}_{4}^{-1}$;
  $\boldsymbol{T_{800}}$: $\mathrm{M}_{1}\mathrm{M}_{4}^{-1}$
} & 1.116
\\
9 & \multicolumn{4}{l}{
  $\boldsymbol{T_{300,400}}$: $\mathrm{M}_{3}\mathrm{M}_{7}\mathrm{M}_{4}\mathrm{M}_{4}^{-1}$;
  $\boldsymbol{T_{800}}$: $\mathrm{M}_{1}\mathrm{M}_{4}^{-1}\mathrm{M}_{7}$
} & 1.116
\\
11 & \multicolumn{4}{l}{
  $\boldsymbol{T_{300}}$: $\mathrm{M}_{3}\mathrm{M}_{7}\mathrm{M}_{4}\mathrm{M}_{4}^{-1}\mathrm{M}_{7}$;
  $\boldsymbol{T_{400,800}}$: $\mathrm{M}_{2}\mathrm{M}_{6}$
} & 1.116/1.392/1.392
\\
12 & \multicolumn{4}{l}{
  $\boldsymbol{T_{300}}$: $\mathrm{M}_{3}\mathrm{M}_{4}\mathrm{M}_{4}^{-1}\mathrm{M}_{3}^{-1}$;
  $\boldsymbol{T_{400}}$: $\mathrm{M}_{3}\mathrm{M}_{5}\mathrm{M}_{4}^{-1}\mathrm{M}_{3}^{-1}$;
  $\boldsymbol{T_{800}}$: $\mathrm{M}_{1}\mathrm{M}_{1}^{-1}$
} & 1.116/1.116/1.061
\\
14 & \multicolumn{4}{l}{
  $\boldsymbol{T_{300}}$: $\mathrm{M}_{3}\mathrm{M}_{7}\mathrm{M}_{4}\mathrm{M}_{4}^{-1}\mathrm{M}_{3}^{-1}$;
  $\boldsymbol{T_{400}}$: $\mathrm{M}_{3}\mathrm{M}_{5}\mathrm{M}_{4}^{-1}\mathrm{M}_{7}\mathrm{M}_{3}^{-1}$;
  $\boldsymbol{T_{800}}$: $\mathrm{M}_{1}\mathrm{M}_{5}^{-1}\mathrm{M}_{7}\mathrm{M}_{3}^{-1}$
} & 1.116/1.116/1.119
\\
18 & \multicolumn{4}{l}{
  $\boldsymbol{T_{300}}$: $\mathrm{M}_{3}\mathrm{M}_{4}\mathrm{M}_{4}^{-1}$;
  $\boldsymbol{T_{400}}$: $\mathrm{M}_{3}\mathrm{M}_{5}\mathrm{M}_{4}^{-1}$;
  $\boldsymbol{T_{800}}$: $\mathrm{M}_{1}\mathrm{M}_{5}^{-1}$
} & 1.116/1.116/1.119
\\
20 & \multicolumn{4}{l}{
  $\boldsymbol{T_{300}}$: $\mathrm{M}_{3}\mathrm{M}_{7}\mathrm{M}_{4}\mathrm{M}_{4}^{-1}$;
  $\boldsymbol{T_{400}}$: $\mathrm{M}_{3}\mathrm{M}_{5}\mathrm{M}_{4}^{-1}\mathrm{M}_{7}$;
  $\boldsymbol{T_{800}}$: $\mathrm{M}_{1}\mathrm{M}_{5}^{-1}\mathrm{M}_{7}$
} & 1.116/1.116/1.119
\\
24 & \multicolumn{4}{l}{
  $\boldsymbol{T_{300}}$: $\mathrm{M}_{4}\mathrm{M}_{4}^{-1}$;
  $\boldsymbol{T_{400,800}}$: $\mathrm{M}_{5}\mathrm{M}_{4}^{-1}$
} & 1.116
\\
26 & \multicolumn{4}{l}{
  $\boldsymbol{T_{300,400}}$: $\mathrm{M}_{7}\mathrm{M}_{4}\mathrm{M}_{4}^{-1}$;
  $\boldsymbol{T_{800}}$: $\mathrm{M}_{7}\mathrm{M}_{5}\mathrm{M}_{4}^{-1}$
} & 1.116
\\
28 & \multicolumn{4}{l}{
  $\boldsymbol{T_{300,400}}$: $\mathrm{M}_{7}\mathrm{M}_{4}\mathrm{M}_{4}^{-1}\mathrm{M}_{7}$;
  $\boldsymbol{T_{800}}$: $\mathrm{M}_{7}\mathrm{M}_{5}\mathrm{M}_{4}^{-1}\mathrm{M}_{7}$
} & 1.116
\\
29 & \multicolumn{4}{l}{
  $\boldsymbol{T_{300}}$: $\mathrm{M}_{4}\mathrm{M}_{4}^{-1}$;
  $\boldsymbol{T_{400}}$: $\mathrm{M}_{5}\mathrm{M}_{4}^{-1}$;
  $\boldsymbol{T_{800}}$: $\mathrm{M}_{5}\mathrm{M}_{5}^{-1}$
} & 1.116/1.116/1.119
\\
31 & \multicolumn{4}{l}{
  $\boldsymbol{T_{300}}$: $\mathrm{M}_{7}\mathrm{M}_{4}\mathrm{M}_{4}^{-1}$;
  $\boldsymbol{T_{400}}$: $\mathrm{M}_{5}\mathrm{M}_{4}^{-1}\mathrm{M}_{7}$;
  $\boldsymbol{T_{800}}$: $\mathrm{M}_{5}\mathrm{M}_{5}^{-1}\mathrm{M}_{7}$
} & 1.116/1.116/1.119
\\
33 & \multicolumn{4}{l}{
  $\boldsymbol{T_{300,400}}$: $\mathrm{M}_{3}^{-1}\mathrm{M}_{2}$;
  $\boldsymbol{T_{800}}$: $\mathrm{M}_{2}^{-1}\mathrm{M}_{3}$
} & 1.392/1.392/0.880
\\
\bottomrule
  \end{tabularx}
\end{table*}

The tetrahedral-site vacancy likewise connects to reconstructed states through distinct routes. Motif $\mathrm{M}_{4}$ contains three transition states and passes through split and three-split intermediates before reaching another split configuration. Motif $\mathrm{M}_{5}$ reaches the corresponding split configuration through a single transition state. These alternatives are particularly useful for understanding pathway competition because they connect comparable endpoint configurations, while differing in the number and energies of intervening minima. Motif $\mathrm{M}_{6}$ connects tetrahedral-site vacancies through a split intermediate, while $\mathrm{M}_{7}$ provides a two-step connection involving displacement out of the Ga-containing plane. Together, these motifs illustrate how local reconstruction accompanies vacancy transport over a range of endpoint separations.

The split $\leftrightarrow$ three-split $\leftrightarrow$ split sequence within $\mathrm{M}_{4}$ is structurally consistent with the three-split-mediated transformation between split-vacancy configurations reported previously.\cite{FrodasonVJV23} The discrete path stationary-point description properly resolves the minima and transition states that make up these multistep migration motifs. This distinction matters because a reconstructed intermediate can introduce further branching even when the net vacancy displacement resembles a previously identified hop. This structural correspondence does not imply quantitative agreement in the barrier heights, because the previous charge-state-specific calculations and the present MLIP landscape are based on different energetic models.

The networks we have sampled also contain four-split vacancies, including the $icbc$-type reconstruction identified by Lee et al.~as an intermediate in the dominant vacancy-diffusion pathway along the $c$ axis.\cite{LeeSE24}  Four-split configurations do not appear along the Dijkstra-selected pathways between point-vacancy endpoints because these pathways are selected based on individual path probabilities for vacancy migration rather than for overall contributions to long-range diffusion. The recovery of the $icbc$-type reconstruction without explicitly prescribing extended vacancy structures nevertheless demonstrates the ability of landscape exploration to identify complex defect configurations beyond those represented by the seven migration motifs.
Lee et al. also showed that increasing the supercell size can substantially lower the relative formation energies of extended three- and four-split vacancies,\cite{LeeSE24} suggesting that finite-size effects may influence their stability in our smaller supercell. We plan to investigate paths for direction-dependent diffusivity and size effects in future work.

The full set of selected pathways is summarised in Table~\ref{tab:pathways} as combinations of these motifs. Repetition of a motif can increase the net displacement without increasing the maximum barrier. For example, paths 25 and 27 contain one and two occurrences of $\mathrm{M}_{7}$, respectively, but both have a maximum barrier of 0.86\,eV. These repeated values reflect the reuse of common local rearrangements, but do not imply equal path probabilities or migration times because longer pathways encounter additional branching points.

\subsection{Temperature-dependent pathway selection}

\begin{figure*}[!t]
  \centering
  \includegraphics[width=0.98\textwidth]{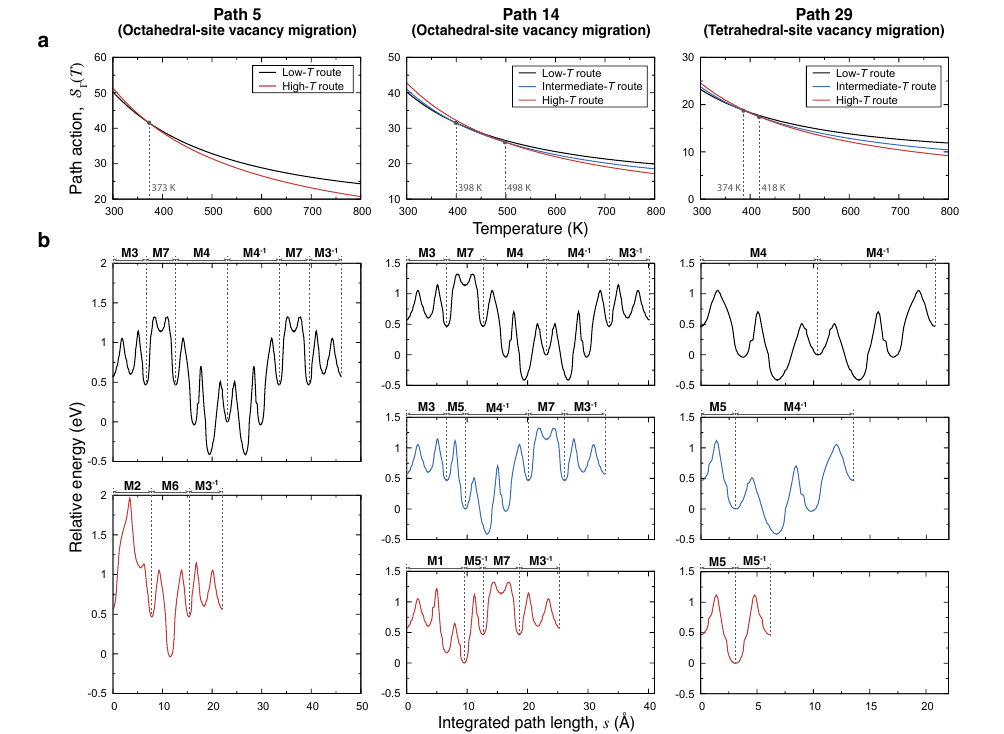}
  \caption{Competition between migration routes for paths 5, 14, and 29. Paths 5 and 14 describe vacancy migration between octahedral Ga sites, whereas path 29 describes migration between tetrahedral Ga sites. (a) Path actions of the competing routes as a function of temperature; the lowest action identifies the most probable sequence among the routes compared. Low-, intermediate-, and high-temperature routes correspond to those selected at 300, 400, and 800\,K, respectively; for path 5, the same route has the highest weight at 400 and 800\,K. Markers indicate crossings at which the preferred route changes. (b) Corresponding potential energy profiles for the competing routes, plotted against the integrated path length, $s$. Energies are referenced to the lowest-energy split-vacancy minimum. Labels above the profiles indicate the sequence of migration motifs $\mathrm{M}_{1}$--$\mathrm{M}_{7}$ defined in Fig.~\ref{fig:pathways} and Table~\ref{tab:pathways}.}
  \label{fig:EofS}
\end{figure*}

Temperature changes the preferred pathways through the kinetic transition network by altering the relative transition rates and hence the branching probabilities at intermediate minima. Of the 34 endpoint pairs, 18 retain the same maximum weight pathway, determined by equation (\ref{eq:weight}), at all three sampled temperatures (300, 400, and 800\,K), while 16 change at least once (Table~\ref{tab:pathways}). Paths 5, 14, and 29 were chosen to illustrate complementary aspects of temperature-dependent route competition. Path 5 illustrates selection of a shorter route despite a higher maximum barrier; path 14 illustrates competition between routes with similar maximum barriers; and path 29 illustrates successive replacement of multistep $\mathrm{M}_{4}$-based segments by more direct $\mathrm{M}_{5}$-based alternatives. Figure~\ref{fig:EofS}(a) compares the actions of the distinct routes selected at the three sampled temperatures; crossings indicate changes in their relative preference. Because the path action depends on the branching probabilities along the entire route, the selection is not determined by the highest energy barrier alone. The corresponding potential energy profiles are shown in Fig.~\ref{fig:EofS}(b). These profiles are independent of temperature, while the branching probabilities change with $T$.

For path 5, the two routes exchange their action ordering near 373\,K. The route favoured at low temperature contains 14 transition states and visits several deep reconstructed minima, whereas the most favourable route at 400 and 800\,K contains only six transition states and has a substantially shorter integrated path length. The latter sequence begins with $\mathrm{M}_{2}$, raising the maximum barrier from 1.12 to 1.39\,eV. For path 29, route selection changes in two stages: the six-step sequence $\mathrm{M}_{4}\mathrm{M}_{4}^{-1}$ is replaced by the four step $\mathrm{M}_{5}\mathrm{M}_{4}^{-1}$ above approximately 374\,K and then by the two step $\mathrm{M}_{5}\mathrm{M}_{5}^{-1}$ above approximately 418\,K. These changes progressively remove excursions through deep intermediate minima. All three routes have similar maximum barriers of approximately 1.12\,eV. The three routes for path 14 provide a complementary case. All have similar maximum barriers of approximately 1.12\,eV, yet their actions cross near 398 and 498\,K, with the lowest action shifting from the low- to the intermediate- and then to the high-temperature route. Together, these examples show that the route with the lower maximum barrier need not remain preferred as the temperature increases, and that changes in the maximum barrier do not track every change in pathway selection. The branching probabilities accumulated over the complete route, including those associated with intermediate minima and additional steps, determine the path action.

These examples distinguish three aspects of vacancy migration: the structural sequence, the energy profile along that sequence, and the path probability within the global network. At each branching point, the branching probability includes competition with all other exits from the same minimum, including exits that do not appear in the displayed profile. Consequently, the profiles alone do not determine the route ordering. At the route level, increasing temperature reduces the Boltzmann penalty associated with higher energy transition states, while shortening a multistep path reduces the number of branching events. Whether a shorter route is more likely therefore depends on the probabilities of the transitions along that route.

The crossings mark changes in the most probable individual sequence, rather than abrupt changes in the underlying landscape or the full ensemble of migration trajectories. Near a crossing, competing routes can have similar probabilities, and routes with lower weights, $\mathcal{W}[\Gamma;T]$, remain accessible. Nor does the most probable sequence necessarily give the shortest mean time to reach the target, since the action does not include the residence times spent in intermediate minima. The network analysis therefore establishes how the preferred structural sequence changes with temperature, while the MFPT analysis below quantifies relaxation into the lowest-energy region. The complete set of potential energy profiles for all 34 endpoint pairs is provided for NequIP-LES and NequIP-OAM-L in Figures~S1--S8, with comparative route statistics reported in Table~S3 of the Supporting Information.

\subsection{Disconnectivity graphs and landscape organisation}

\begin{figure}[!t]
  \centering
  \includegraphics[width=0.8\columnwidth]{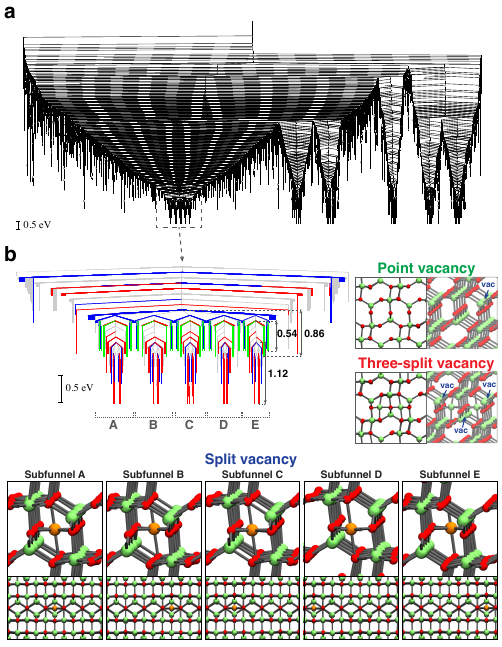}
  \caption{Disconnectivity graph of the sampled vacancy landscape: (a) full graph and (b) enlargement of the boxed low-energy region. The upper energy limits are 13.54 and 2.54\,eV above the lowest-energy state for the full graph and enlargement, respectively, with energy levels for basin merging spaced by 0.1\,eV. Vertical scale bars denote 0.5\,eV. In the enlargement, green, blue, and red identify point, split, and three-split vacancies, respectively; other branches are grey. Subfunnels A--E contain symmetry-related vacancy configurations in different Ga-containing planes. Representative point and three-split structures illustrate the structural assignments, while split-vacancy examples are shown below for each subfunnel in two views. Orange spheres highlight the Ga atom involved in the split-vacancy reconstruction.}
  \label{fig:DG}
\end{figure}

The disconnectivity graph in Fig.~\ref{fig:DG}(a) provides a global view of the sampled vacancy landscape and contains several major funnels. We refer to the broad leftmost funnel as the main funnel, and the boxed low-energy region is enlarged in Fig.~\ref{fig:DG}(b). In this study, the symmetry-related minima were retained as distinct entries in both the database and the graph to resolve migration pathways connecting equivalent configurations at different lattice positions. Database expansion focused on the 34 migration pathways within the main funnel, which was sampled sufficiently for the present study. Consequently, the detailed branching structures of the other symmetry-related funnels were not explored to the same extent and may be less well converged. Branches within the main funnel merge at different energy thresholds, revealing a hierarchy of subbasins such that relaxation need not proceed through a single sequence of rearrangements.

Figure~\ref{fig:DG}(b) shows five symmetry-related subfunnels, A--E, associated with different Ga-containing planes, as illustrated by the split-vacancy structures below the enlargement. The fivefold repetition reflects equivalent vacancy reconstructions at different positions in the chosen supercell, rather than five distinct defect structures. Green, blue, and red identify point, split, and three-split vacancies, respectively; other branches are grey. Three-split structures occupy the deepest branches, consistent with the energetic preference discussed above, while split and point vacancies occur at higher energies within the surrounding subbasins. The energy differences highlighted in Fig.~\ref{fig:DG}(b) correspond to selected local barriers along the 300\,K route of path 5. Starting from an octahedral-site point vacancy, the route reaches a tetrahedral-site point vacancy through $\mathrm{M}_{3}$, with a maximum local barrier of 0.54\,eV. It then follows $\mathrm{M}_{7}$ to an equivalent vacancy configuration in a different Ga-containing plane, crossing a local barrier of 0.86\,eV and connecting different subfunnels. The route passes through low-energy three-split intermediates and encounters its largest local barrier, 1.12\,eV, upon escape through $\mathrm{M}_{4}^{-1}$.

This organisation provides a network context for the recurring migration motifs. Vacancy migration connects minima within a landscape that contains many alternative intermediates, and similar local configurations can be reached through different sequences. The graph reveals the energy thresholds required to connect these regions, but does not specify how rapidly they exchange population. In particular, the horizontal separation of branches carries no information about atomic displacement, and a low-lying minimum need not have a short relaxation time to the lowest-energy set. The kinetic consequences of this hierarchy of barriers are quantified in the next section using MFPT analysis.

\subsection{Mean first-passage times to low energy states}

\begin{figure}[!t]
  \centering
  \includegraphics[width=0.85\columnwidth]{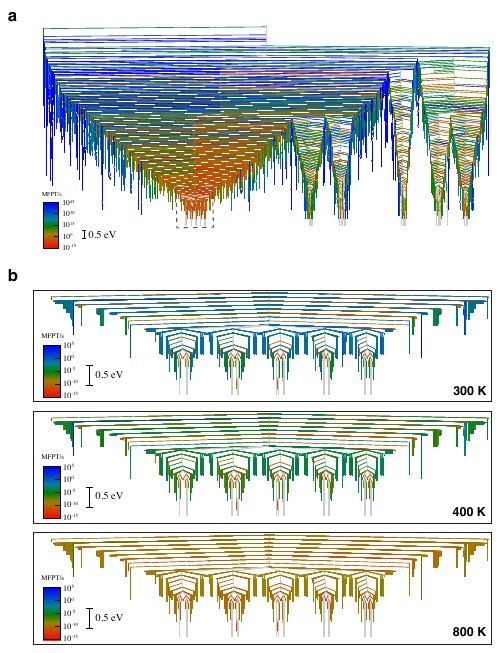}
  \caption{Disconnectivity graphs coloured by the mean first-passage time (MFPT) from each minimum to the set of 48 lowest-energy minima. The 48 target minima are shown in grey. (a) Full landscape at 300\,K. (b) Enlargements of the low-energy region at 300, 400, and 800\,K. Colours represent the MFPT in seconds on logarithmic scales; panels (a) and (b) use different ranges, while all three graphs in (b) share one scale. Values above the upper limit in panel (a) are coloured by the maximum value. Vertical scale bars denote 0.5\,eV.}
  \label{fig:MFPT}
\end{figure}

Figure~\ref{fig:MFPT} shows the mean time required to reach any of the 48 minima in the lowest-energy set, shown in grey, starting from each of the other minima. The target minima are absorbing states and therefore have zero MFPT by definition. These target minima are chosen here as the three-split configurations belonging to the lowest-energy family, so the MFPTs describe relaxation to this reconstructed vacancy set. At 300\,K, the full graph exhibits a wide variation in MFPT across the landscape. Some branches connect rapidly to the target region, whereas other groups retain much longer first-passage times. The contrast between these groups shows that energetic stability alone does not determine access to the lowest-energy states: minima in a separate basin may require slow escape events before reaching the target set. The longest MFPTs may be particularly sensitive to incompletely sampled escape pathways and should therefore not be regarded as well-converged physical predictions. The full graph is retained for context, while the discussion focuses on the low-energy region examined in the migration analysis.

The magnified graphs use a common colour scale to reveal the temperature dependence within the low-energy region. For the 24 point-vacancy endpoint configurations defined in the 34 migration pairs, the MFPTs are of the order of $10^{-1}$\,s at 300\,K, $10^{-4}$\,s at 400\,K, and $10^{-9}$\,s at 800\,K. At each temperature, the values for these endpoints differ by no more than a factor of approximately 2.5, whereas the characteristic times decrease by nearly eight orders of magnitude between 300 and 800\,K. The effect of temperature is therefore substantially greater than the variation among the point-vacancy endpoint configurations. The similar timescales, despite the different migration sequences, indicate that the choice of point-vacancy starting configuration has a relatively small effect on the overall relaxation time. The underlying PES is unchanged; the changing colours instead reflect the temperature dependence of the transition rates through the network. We note again here that the rates and time scales are approximate. However, we expect the key trends that we identify to be generally reliable when the differences are large.

These MFPTs complement the route comparisons in Fig.~\ref{fig:EofS}. A selected route describes one sequence of transitions, whereas the MFPT includes residence times, all possible routes, and repeated visits to intermediate states before arrival. A compact route can therefore coexist with slow relaxation if an intermediate is difficult to escape from, while several accessible routes can facilitate arrival without any one sequence dominating. The MFPT thus provides a network-level measure of relaxation that cannot be inferred from any single selected pathway. These MFPTs quantify the time required for a sampled Ga-vacancy configuration to first reach any member of the lowest-energy three-split set. They therefore exclude vacancy formation and subsequent equilibration within the target set and should not be interpreted as long-range vacancy-diffusion times. In the absence of a directly comparable experimental measurement, the absolute values should be regarded as model-dependent harmonic transition state theory estimates.

\subsection{Heat capacity and equilibrium populations}

\begin{figure}[!t]
\centering
\includegraphics[width=0.8\columnwidth]{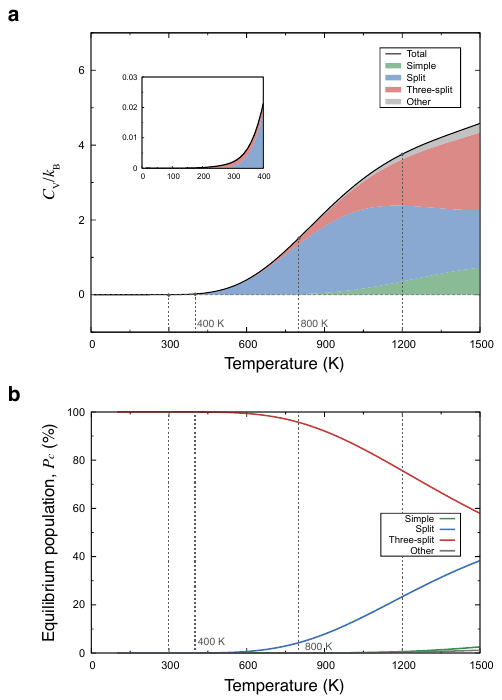}
\caption{Equilibrium thermodynamics of the sampled landscape within the classical harmonic superposition approximation. (a) Excess heat capacity referenced to the common harmonic baseline and resolved by structural class. The inset magnifies the low temperature response. (b) Equilibrium populations of the structural classes. Green, blue, red, and grey denote point, split, three-split, and other configurations, respectively, in panels (a) and (b).}
\label{fig:Cv}
\end{figure}

The excess heat capacity in Fig.~\ref{fig:Cv}(a), referenced to the common harmonic baseline, is close to zero at low temperature and increases as the temperature rises. The excess contribution is very small at 300 and 400\,K, as shown in the inset, but becomes progressively larger as higher energy minima acquire greater statistical weight. No distinct peak is observed within the plotted temperature range; instead, the excess contribution increases smoothly with temperature. This behaviour is consistent with a gradual thermal redistribution across the sampled landscape, rather than a sharp thermodynamic anomaly within this temperature range.

The equilibrium populations in Fig.~\ref{fig:Cv}(b) show this redistribution more directly. Three-split vacancies remain strongly favoured over a broad temperature range, accounting for more than 99.9\% of the population at 400\,K and approximately 95.7\% at 800\,K. Split vacancies form the main competing population, reaching about 4.3\% at 800\,K, whereas point vacancies remain sparsely populated. At the upper end of the plotted range, 1500\,K, the three-split population has fallen to roughly 58\% and the split population has risen to about 38\%. Three-split configurations therefore remain the most populated structural family throughout the range shown. However, their dominance in the equilibrium population, does not translate directly into a heat capacity signature.

The structural class decomposition shows that equilibrium abundance alone does not determine the heat capacity response. At 800\,K, split vacancies account for only a few percent of the equilibrium population, yet contribute approximately 88\% of the excess heat capacity. In contrast, the overwhelmingly more populated three-split configurations make a much smaller contribution to the excess at this temperature. This difference arises because the excess heat capacity reflects energy fluctuations rather than population alone. As temperature increases, population is transferred from the low-energy three-split states to higher-energy split configurations, so even a relatively small split population can make a large contribution to the energy variance. The coloured areas should accordingly be interpreted as contributions to the total energy variance, rather than as independent heat capacities of the individual structural classes. Disconnectivity graphs highlighting the minima that gain or lose equilibrium population on heating are provided in Fig.~S9 of the Supporting Information.

The population curves also put the temperature dependent route changes into perspective. Around 400\,K, where some of the competing migration routes change their ordering, the equilibrium ensemble remains almost entirely three-split. A change in the most probable migration sequence can thus occur without a substantial change in the dominant vacancy family. This difference arises because route selection reflects competition among the available transitions at each intermediate minimum, whereas equilibrium populations depend on the statistical weights of its minima. At the upper end of the temperature range, these results should be interpreted as qualitative trends, because the harmonic approximation does not account for local vibrational anharmonic effects or configurations outside the sampled stationary point database. Together, these results show how a strong equilibrium preference for reconstructed vacancies can coexist with a variety of migration sequences and widely differing relaxation times.

\section{Conclusion}

We have developed a persistent socket interface that connects the GMIN and OPTIM programs to ASE-compatible calculators and supports energy landscape exploration coordinated by PATHSAMPLE. This coupling facilitates global optimisation and transition state searches using atomistic models available through ASE. Using this interface, we investigated Ga vacancies in $\beta$-Ga$_{2}$O$_{3}$, bringing structure searches, migration pathways, relaxation kinetics, and equilibrium thermodynamics together through a common stationary point database.

Vacancy reconstruction produces low-energy split and three-split configurations that additionally serve as intermediates along migration pathways. Connections between 34 symmetry-inequivalent vacancy endpoint pairs can be described in terms of seven recurring structural motifs, several of which contain multiple minima and transition states. As temperature varies, changes in the branching probabilities of outgoing transitions at intermediate minima alter the relative probabilities of complete migration sequences. Consequently, the route corresponding to the highest probability measure, $\mathcal{W}[\Gamma;T]$, cannot be inferred from the highest energy barrier alone. The disconnectivity graphs reveal a hierarchical landscape containing symmetry-related funnels, while the MFPT analysis quantifies relaxation to the lowest-energy regions of the sampled network. Understanding the complex kinetics associated with alternative pathways requires a detailed consideration of the associated transition network.

The thermodynamic analysis provides a complementary perspective. Three-split vacancies retain the largest equilibrium population up to 1500\,K, while the heat-capacity analysis by vacancy type shows that the less populated split configurations can make a substantial contribution to the excess heat capacity as higher-energy states become populated. This redistribution is distinct from the kinetic route changes: the preferred migration route can change while the identity of the dominant equilibrium vacancy family remains the same. Together, these results distinguish the configurations favoured at equilibrium from the pathways and timescales through which the defect relaxes.

Understanding these reconstructions is important for modelling atomic transport and dopant redistribution in $\beta$-Ga$_2$O$_3$. Split Ga vacancies in $\beta$-Ga$_2$O$_3$ were originally predicted computationally,\cite{varley_hydrogenated_2011} before experimental confirmation with positron annihilation spectroscopy\cite{KarjalainenMEG21} and hydrogen-related vibrational spectroscopy.\cite{Weiser18} 
Previous calculations have shown that three-split intermediates can lower vacancy migration barriers\cite{FrodasonVJV23} and that networks of reconstructed vacancies can be used to calculate directional diffusivities.\cite{LeeSE24} Both theory and experiment have shown that Sn diffusion, controlling $n$-type doping in Ga$_2$O$_3$, can be mediated by Ga vacancies, further demonstrating the relevance of these defects to dopant redistribution.\cite{FrodasonKVV23,RockYEL24} By resolving reconstructed vacancy configurations, their migration motifs, and their temperature-dependent kinetics, the present work connects these transport-relevant defect structures directly to their migration behaviour. This combined description shows how the same reconstructed configurations can influence both the available migration pathways and the timescales for relaxation.

More generally, the risk of missing important defect configurations through relaxation of idealised starting structures has been emphasised in previous studies of defect energy landscapes.\cite{MosqueraK21,MosqueraKWS23} 
Knowledge of defect energy landscapes is crucial for accurate defect modelling, from formation energies and thermodynamics, to migration and defect-related processes such as radiative or non-radiative carrier recombination.
Chemically guided bond distortions and rattling, as implemented in ShakeNBreak, provide an efficient route to energy-lowering reconstructions, but more non-local rearrangements such as split-vacancy formation present a more challenging configurational search problem.\cite{MosqueraKWS23,Kavanagh25} In $\beta$-Ga$_2$O$_3$, split-vacancy structures have previously been identified (serendipitously) via NEB calculations of point-vacancy migration or through explicit generation and screening of candidate split vacancy geometries.\cite{Kavanagh25} Here, basin-hopping searches starting only from conventional point-vacancy configurations recovered both split and three-split vacancies without knowledge of these reconstructed geometries or explicitly enumerating candidate split-vacancy structures. Subsequent transition state searches and discrete path sampling then connected these configurations into a common network for kinetic and thermodynamic analysis. The ASE integration into a full potential energy landscape exploration framework provides a practical basis for systematic prediction and understanding of complex defect landscapes and can be employed for any other  systems with complex structural rearrangements accessible via ASE-compatible atomistic models.

\section*{Supporting Information}
Bytewise verification of response-data integrity and computational timing of the socket interface; potential-energy profiles of the Dijkstra routes selected at approximately 300, 400, and 800\,K for all 34 symmetry-inequivalent endpoint pairs obtained using NequIP-LES and NequIP-OAM-L; comparison of the stationary-point database sizes, temperature-dependent route selection, mean numbers of elementary steps, and mean maximum barriers for the two potentials; vacancy-class counts before and after symmetry grouping, including point, split, three-split, four-split, and other configurations; and disconnectivity graphs highlighting thermal population redistribution (PDF).

\section*{Acknowledgments}
S.W.P. and D.J.W. gratefully acknowledge the support of Xyme.

\section*{Author Declarations}
\subsection*{Conflict of Interest}
The authors have no conflicts to disclose.

\section*{Data Availability Statement}
The Cambridge Energy Landscapes software is available for download from https://www-wales.ch.cam.ac.uk/software.html. The data supporting the findings of this study are available on Zenodo at https://doi.org/10.5281/zenodo.23271460.

\bibliographystyle{unsrtnat}
\bibliography{myrefs,SK_refs}

@article{mosquera-lois_search_2021,
	title = {In search of hidden defects},
	volume = {4},
	issn = {2590-2393, 2590-2385},
	url = {https://www.cell.com/matter/abstract/S2590-2385(21)00273-3},
	doi = {10.1016/j.matt.2021.06.003},
	language = {English},
	number = {8},
	urldate = {2021-08-17},
	journal = {Matter},
	publisher = {Elsevier},
	author = {Mosquera-Lois, Irea and Kavanagh, Seán R.},
	month = aug,
	year = {2021},
	pages = {2602--2605},
}

@article{varley_hydrogenated_2011,
	title = {Hydrogenated cation vacancies in semiconducting oxides},
	volume = {23},
	issn = {0953-8984, 1361-648X},
	url = {https://iopscience.iop.org/article/10.1088/0953-8984/23/33/334212},
	doi = {10.1088/0953-8984/23/33/334212},
	language = {en},
	number = {33},
	urldate = {2023-09-20},
	journal = {Journal of Physics: Condensed Matter},
	author = {Varley, J B and Peelaers, H and Janotti, A and Van De Walle, C G},
	month = aug,
	year = {2011},
	pages = {334212},
}

@article{mosquera-lois_identifying_2023,
	title = {Identifying the ground state structures of point defects in solids},
	volume = {9},
	copyright = {2023 The Author(s)},
	issn = {2057-3960},
	url = {https://www.nature.com/articles/s41524-023-00973-1},
	doi = {10.1038/s41524-023-00973-1},
	language = {en},
	number = {1},
	urldate = {2023-02-18},
	journal = {npj Computational Materials},
	publisher = {Nature Publishing Group},
	author = {Mosquera-Lois, Irea and Kavanagh, Seán R. and Walsh, Aron and Scanlon, David O.},
	month = feb,
	year = {2023},
	note = {Number: 1},
	pages = {1--11},
}

@article{lis87,
  author    = {Li, Z. and Scheraga, H. A.},
  title     = {Monte Carlo-minimization approach to the multiple-minima problem in protein folding},
  journal   = {Proc. Natl. Acad. Sci. {USA}},
  pages     = {6611-6615},
  year      = {1987},
  volume    = {84}
}

@article{WalesS99,
  title={Global optimization of clusters, crystals, and biomolecules},
  author={Wales, David J and Scheraga, Harold A},
  journal={Science},
  volume={285},
  number={5432},
  pages={1368--1372},
  year={1999},
  publisher={American Association for the Advancement of Science}
}

@article{RockYEL24,
  title={Utilizing {(Al, Ga)$_2$O$_3$/Ga$_2$O$_3$} superlattices to measure cation vacancy diffusion and vacancy-concentration-dependent diffusion of {Al}, {Sn}, and {Fe} in {$\beta$-Ga$_2$O$_3$}},
  author={Rock, Nathan D and Yang, Haobo and Eisner, Brian and Levin, Aviva and Bhattacharyya, Arkka and Krishnamoorthy, Sriram and Ranga, Praneeth and Walker, Michael A and Wang, Larry and Cheng, Ming Kit and others},
  journal={APL Materials},
  volume={12},
  number={8},
  year={2024},
  publisher={AIP Publishing}
}

@article{FrodasonKVV23,
  title={Diffusion of {Sn} donors in {$\beta$-Ga$_2$O$_3$}},
  author={Frodason, Ymir K and Krzyzaniak, Patryk P and Vines, Lasse and Varley, Joel B and Van de Walle, Chris G and Johansen, Klaus Magnus H},
  journal={APL Materials},
  volume={11},
  number={4},
  year={2023},
  publisher={AIP Publishing}
}

@article{MosqueraKWS23,
  title={Identifying the ground state structures of point defects in solids},
  author={Mosquera-Lois, Irea and Kavanagh, Se{\'a}n R and Walsh, Aron and Scanlon, David O},
  journal={npj Computational Materials},
  volume={9},
  number={1},
  pages={25},
  year={2023},
  publisher={Nature Publishing Group UK London}
}

@article{MosqueraK21,
  title={In search of hidden defects},
  author={Mosquera-Lois, Irea and Kavanagh, Se{\'a}n R},
  journal={Matter},
  volume={4},
  number={8},
  pages={2602--2605},
  year={2021},
  publisher={Elsevier}
}

@article{Weiser18,
  title={Structure and vibrational properties of the dominant OH center in {$\beta$-Ga$_2$O$_3$}},
  author={Weiser, Philip and Stavola, Michael and Fowler, W Beall and Qin, Ying and Pearton, Stephen},
  journal={Applied Physics Letters},
  volume={112},
  number={23},
  year={2018},
  publisher={AIP Publishing}
}

@article{SchonPJ96,
  title={Studying the energy hypersurface of continuous systems-the threshold algorithm},
  author={Sch{\"o}n, JC and Putz, H and Jansen, M},
  journal={Journal of Physics: Condensed Matter},
  volume={8},
  pages={143--156},
  year={1996}
}

@article{Schon24,
  title={Energy landscapes—Past, present, and future: A perspective},
  author={Sch{\"o}n, JC},
  journal={The Journal of Chemical Physics},
  volume={161},
  number={5},
  year={2024},
  publisher={AIP Publishing}
}

@article{Heuer97,
  title={Properties of a glass-forming system as derived from its potential energy landscape},
  author={Heuer, Andreas},
  journal={Physical review letters},
  volume={78},
  pages={4051},
  year={1997}
}

@article{mcginty71,
  author    = {D. J. McGinty},
  title     = {Vapor phase homogenous nucleation and the thermodynamic properties of small clusters
              of argon atoms},
  journal   = {J. Chem. Phys.},
  pages     = {580},
  year      = {1971},
  volume    = {55}
}

@article{burton72,
  author    = {J. J. Burton},
  title     = {Vibrational Frequencies and Entropies of Small Clusters of Atoms},
  journal   = {J. Chem. Phys.},
  pages     = {3133},
  year      = {1972},
  volume    = {56}
}

@article{hoare79,
  author    = {M. R. Hoare},
  title     = {Structure and dynamics of simple microclusters},
  journal   = {Adv. Chem. Phys.},
  pages     = {49-135},
  year      = {1979},
  volume    = {40}
}

@article{stillingerw84,
  author    = {Stillinger, F. H.  and Weber, T. A.},
  title     = {packing structures and transitions in liquids and solids},
  journal   = {Science},
  pages     = {983-989}, year      = {1984}, volume    = {225}
}

@article{frankehb93,
  title     = {The structure of small clusters: multiple normal modes
              model},
  author    = {Franke, G. and Hilf, E. R. and Borrmann, P.},
  journal   = {J. Chem. Phys.},
  pages     = {3496},
  year      = {1993},
  volume    = {98}
}

@article{henkelmanj99,
  author    = {Henkelman,  G. and J\'onsson,  H.},
  title     = {A dimer method for finding saddle points on high
              dimensional potential surfaces using only first
              derivatives},
  journal   = {J. Chem. Phys.},
  year      = {1999},
  volume    = {111},
  pages     = {7010-7022}
}

@article{ZengXH14,
   author = "Zeng, Yi and Xiao, Penghao and Henkelman, Graeme",
   title = "Unification of algorithms for minimum mode optimization",
   journal = "J. Chem. Phys.",
   year = "2014",
   volume = "140",
   number = "4",
   eid = 044115,
   pages = "044115",
   doi = "http://dx.doi.org/10.1063/1.4862410"
}

@article{SharpeW21c,
author = {Sharpe,Daniel J.  and Wales,David J. },
title = {Nearly reducible finite Markov chains: Theory and algorithms},
journal = {J. Chem. Phys.},
volume = {155},
pages = {140901},
year = {2021},
}

@Article{WoodsW24,
author ="Woods, Esmae J. and Wales, David J.",
title  ="Analysis and interpretation of first passage time distributions featuring rare events",
journal  ="Phys. Chem. Chem. Phys.",
year  ="2024",
volume  ="26",
pages  ="1640-1657",
doi  ="10.1039/D3CP04199A",
}

@article{PerdewRCV08,
  title={Restoring the density-gradient expansion for exchange in solids and surfaces},
  author={Perdew, John P and Ruzsinszky, Adrienn and Csonka, G{\'a}bor I and Vydrov, Oleg A and Scuseria, Gustavo E and Constantin, Lucian A and Zhou, Xiaolan and Burke, Kieron},
  journal={Physical review letters},
  volume={100},
  number={13},
  pages={136406},
  year={2008},
  publisher={APS}
}

@article{KresseJ99,
  title={From ultrasoft pseudopotentials to the projector augmented-wave method},
  author={Kresse, Georg and Joubert, Daniel},
  journal={Physical review b},
  volume={59},
  number={3},
  pages={1758},
  year={1999},
  publisher={APS}
}

@article{KresseF96,
  title={Efficient iterative schemes for ab initio total-energy calculations using a plane-wave basis set},
  author={Kresse, Georg and Furthm{\"u}ller, J{\"u}rgen},
  journal={Physical review B},
  volume={54},
  number={16},
  pages={11169},
  year={1996},
  publisher={APS}
}

@article{OngRJH13,
  title={{Python} Materials Genomics ({pymatgen}): A robust, open-source python library for materials analysis},
  author={Ong, Shyue Ping and Richards, William Davidson and Jain, Anubhav and Hautier, Geoffroy and Kocher, Michael and Cholia, Shreyas and Gunter, Dan and Chevrier, Vincent L and Persson, Kristin A and Ceder, Gerbrand},
  journal={Computational Materials Science},
  volume={68},
  pages={314--319},
  year={2013},
  publisher={Elsevier}
}

@article{walesmw98,
  author    = {Wales, D. J. and Miller, M. A. and Walsh, T. R},
  title     = {Archetypal energy landscapes},
  journal   = {Nature},
  year      = {1998},
  volume    = {394},
  pages     = {758-760},
}

@article{beckerk97,
  author    = {Becker, O. M. and Karplus, M.},
  title     = {The topology of multidimensional potential energy surfaces: Theory and application to peptide structure and kinetics},
  journal   = {J. Chem. Phys.},
  pages     = {1495-1517},
  year      = {1997},
  volume    = {106},
  category  = {proteins; master equation},
  comments  = {Master equation. Super basins. Disconnectivity graphs.},
  copied    = {3.iv.97}
}

@article{Broyden70,
title={The Convergence of a Class of Double-rank Minimization Algorithms 1. General Considerations},
author = {Broyden, C. G.}, journal = {J. Inst. Math. Appl.}, volume = {6}, pages = {76-90}, year = {1970} }

@article{Fletcher70,
title={A new approach to variable metric algorithms},
author = {Fletcher, R.}, journal = {Comput. J.}, volume = {13}, pages = {317-322}, year = {1970} }

@article{Goldfarb70,
title={A family of variable-metric methods derived by variational means},
author = {Goldfarb, D.}, journal = {Math. Comput.}, volume = {24}, pages = {23-26}, year = {1970} }

@article{Shanno70,
title={Conditioning of quasi-Newton methods for function minimization},
author = {Shanno, D. F.}, journal = {Math. Comput.}, volume = {24}, pages = {647-656}, year = {1970} }

@article{LiuN89,
author = {Liu, D. and Nocedal, J.},
title = {On the limited memory {BFGS} method for large scale optimization},
journal = {Math. Prog.},
volume = {45},
year = {1989},
pages = {503-528}
}

@article{Wales17,
  title={Decoding heat capacity features from the energy landscape},
  author={Wales, David J},
  journal={Physical Review E},
  volume={95},
  number={3},
  pages={030105},
  year={2017},
  publisher={APS}
}

@article{SharpeW21b,
  title={Numerical analysis of first-passage processes in finite Markov chains exhibiting metastability},
  author={Sharpe, Daniel J and Wales, David J},
  journal={Physical Review E},
  volume={104},
  number={1},
  pages={015301},
  year={2021},
  publisher={APS}
}

@article{SharpeW21a,
  title={Graph transformation and shortest paths algorithms for finite Markov chains},
  author={Sharpe, Daniel J and Wales, David J},
  journal={Physical Review E},
  volume={103},
  number={6},
  pages={063306},
  year={2021},
  publisher={APS}
}

@article{Wales09,
  title={Calculating rate constants and committor probabilities for transition networks by graph transformation},
  author={Wales, David J},
  journal={The Journal of chemical physics},
  volume={130},
  number={20},
  year={2009},
  publisher={AIP Publishing}
}

@article{Wales06,
  title={Energy landscapes: calculating pathways and rates},
  author={Wales, David J},
  journal={International Reviews in Physical Chemistry},
  volume={25},
  number={1-2},
  pages={237--282},
  year={2006},
  publisher={Taylor \& Francis}
}

@article{KavanaghSNM24,
  title={{doped}: {Python} toolkit for robust and repeatable charged defect supercell calculations},
  author={Kavanagh, Se{\'a}n R and Squires, Alexander G and Nicolson, Adair and Mosquera-Lois, Irea and Ganose, Alex M and Zhu, Bonan and Brlec, Katarina and Walsh, Aron and Scanlon, David O},
  journal={Journal of Open Source Software},
  volume={9},
  number={96},
  pages={6433},
  year={2024},
  doi={10.21105/joss.06433}
}

@article{Cheng25,
  title={Latent Ewald summation for machine learning of long-range interactions},
  author={Cheng, Bingqing},
  journal={npj Computational Materials},
  volume={11},
  number={1},
  pages={80},
  year={2025},
  publisher={Nature Publishing Group UK London}
}

@article{KimWVZ25,
  title = {A Universal Augmentation Framework for Long-Range Electrostatics in Machine Learning Interatomic Potentials},
  author = {Kim, Dongjin and Wang, Xiaoyu and Vargas, Santiago and Zhong, Peichen and King, Daniel S. and Inizan, Theo Jaffrelot and Cheng, Bingqing},
  year = 2025,
  journal = {Journal of Chemical Theory and Computation},
  publisher = {American Chemical Society},
  doi = {10.1021/acs.jctc.5c01400}
}

@article{TanDKM26,
  title={High-performance training and inference for deep equivariant interatomic potentials},
  author={Tan, Chuin Wei and Descoteaux, Marc L and Kotak, Mit and de Miranda Nascimento, Gabriel and Kavanagh, Se{\'a}n R and Zichi, Laura and Wang, Menghang and Saluja, Aadit and Hu, Yizhong R and Smidt, Tess and others},
  journal={Digital Discovery},
  volume={5},
  number={4},
  pages={1558--1567},
  year={2026},
  publisher={The Royal Society of Chemistry}
}

@article{WalesD97,
  title={Global optimization by basin-hopping and the lowest energy structures of Lennard-Jones clusters containing up to 110 atoms},
  author={Wales, David J and Doye, Jonathan PK},
  journal={The Journal of Physical Chemistry A},
  volume={101},
  number={28},
  pages={5111--5116},
  year={1997},
  publisher={ACS Publications}
}

@article{BatznerMSG22,
  title={{E(3)}-equivariant graph neural networks for data-efficient and accurate interatomic potentials},
  author={Batzner, Simon and Musaelian, Albert and Sun, Lixin and Geiger, Mario and Mailoa, Jonathan P and Kornbluth, Mordechai and Molinari, Nicola and Smidt, Tess E and Kozinsky, Boris},
  journal={Nature communications},
  volume={13},
  number={1},
  pages={2453},
  year={2022},
  publisher={Nature Publishing Group UK London}
}

@article{BehlerP07,
  title={Generalized neural-network representation of high-dimensional potential-energy surfaces},
  author={Behler, J{\"o}rg and Parrinello, Michele},
  journal={Physical review letters},
  volume={98},
  number={14},
  pages={146401},
  year={2007},
  publisher={APS}
}

@article{KavanaghTWD26,
  title={Fast and Accurate Foundation Models for Equivariant Machine-Learned Interatomic Potentials},
  author={Kavanagh, Se{\'a}n R and Tan, Chuin Wei and Wang, Menghang and Descoteaux, Marc L and Nascimento, Gabriel de Miranda and Unneberg, Ulrik and Zichi, Laura and Libbi, Francesco and Rivano, Norma and Glover, Austin and others},
  journal={arXiv preprint arXiv:2607.28461},
  year={2026}
}

@article{LeeSE24,
  title={Investigation of {Ga} interstitial and vacancy diffusion in {$\beta$-Ga$_2$O$_3$} via split defects: A direct approach via master diffusion equations},
  author={Lee, Channyung and Scarpulla, Michael A and Ertekin, Elif},
  journal={Physical Review Materials},
  volume={8},
  number={5},
  pages={054603},
  year={2024},
  publisher={APS}
}

@article{KarjalainenMEG21,
  title={Split {Ga} vacancies in n-type and semi-insulating {$\beta$-Ga$_2$O$_3$} single crystals},
  author={Karjalainen, A and Makkonen, I and Etula, J and Goto, K and Murakami, H and Kumagai, Y and Tuomisto, F},
  journal={Applied Physics Letters},
  volume={118},
  number={7},
  year={2021},
  publisher={AIP Publishing}
}

@article{Kavanagh25,
  title={Identifying split vacancy defects with machine-learned foundation models and electrostatics},
  author={Kavanagh, Se{\'a}n R},
  journal={Journal of Physics: Energy},
  volume={7},
  number={4},
  pages={045002},
  year={2025},
  publisher={IOP Publishing}
}

@article{HjorthLarsenJBC17,
  title={The atomic simulation environment—a {Python} library for working with atoms},
  author={Larsen, Ask Hjorth and Mortensen, Jens J{\o}rgen and Blomqvist, Jakob and Castelli, Ivano E and Christensen, Rune and Du{\l}ak, Marcin and Friis, Jesper and Groves, Michael N and Hammer, Bj{\o}rk and Hargus, Cory and others},
  journal={Journal of Physics: Condensed Matter},
  volume={29},
  number={27},
  pages={273002},
  year={2017},
  publisher={IOP Publishing}
}

@misc{Pracht26,
  author       = {Pracht, Philipp},
  title        = {fmlip-relay},
  year         = {2026},
  url          = {https://github.com/pprcht/fmlip-relay},
  note         = {Accessed 18 March 2026}
}

@misc{GMIN,
  title   = {{\tt GMIN}: A program for basin-hopping global optimisation, basin-sampling, and parallel tempering},
  address = {http://www-wales.ch.cam.ac.uk/software.html},
  howpublished = {http://www-wales.ch.cam.ac.uk/software.html}
}

@misc{OPTIM,
  title   = {{\tt OPTIM}: A program for geometry optimisation and pathway calculations},
  address = {http://www-wales.ch.cam.ac.uk/software.html},
  howpublished = {http://www-wales.ch.cam.ac.uk/software.html}
}

@misc{PATHSAMPLE,
  title   = {{\tt PATHSAMPLE}: A program for generating connected stationary point databases and
             extracting global kinetics},
  address = {http://www-wales.ch.cam.ac.uk/software.html},
  howpublished = {http://www-wales.ch.cam.ac.uk/software.html}
}

@book{Wales03,
author = {Wales, D. J.}, 
title = {Energy Landscapes}, 
publisher = {Cambridge University Press}, 
address = {Cambridge}, 
year = {2003}
}

@article{TrygubenkoW04,
title = {A Doubly Nudged Elastic Band Method for Finding Transition States},
author = {Trygubenko, S. A. and Wales, D. J.},
journal = {J. Chem. Phys.}, volume = {120}, pages = {2082-2094}, year = {2004} }

@article{SheppardTH08,
author = {Sheppard, D. and Terrell, R. and Henkelman, G.},
journal = {J. Chem. Phys.},
pages = {134106},
title = {{Optimization methods for finding minimum energy paths}},
volume = {128},
year = {2008}
}

@article{Millsjs95,
title = {Reversible work transition state theory: application to dissociative adsorption of hydrogen},
journal = {Surface Science},
volume = {324},
number = {2-3},
pages = {305-337},
year = {1995},
issn = {0039-6028},
author = {Mills, G. and J{\'o}nsson, H. and Schenter, G. K.}}

@InCollection{JonssonMJ98,
  author = {J{\'o}nsson, H. and Mills, G. and Jacobsen, K. W.},
  booktitle = {Classical and Quantum Dynamics in Condensed Phase Simulations},
  editor = {B. J. Berne and G. Ciccotti and D. F. Coker},
  year = {1998},
  pages = {385-404},
  chapter   = {16},
  title     = {Nudged elastic band method for finding minimum energy paths of transitions},
  publisher = {World Scientific},
  address = {Singapore}
}

@ARTICLE{HenkelmanUJ00,
  author    = {Henkelman, G. and Uberuaga, B. P. and J\'onsson, H.},
  title     = {A climbing image nudged elastic band method for finding saddle
              points and minimum energy paths},
  journal   = {J. Chem. Phys.},
  year      = {2000},
  volume    = {113},
  pages     = {9901-9904}
}

@ARTICLE{HenkelmanJ00,
  author    = {Henkelman, G. and J\'onsson, H.},
  title     = {Improved tangent estimate in the nudged elastic band method for finding minimum energy paths and saddle points}, 
  journal   = {J. Chem. Phys.},
  year      = {2000},
  volume    = {113},
  pages     = {9978-9985}
}

@article{munrow99,
  author    = {Munro, L. J. and Wales, D. J.},
  title     = {Defect migration in crystalline silicon},
  journal   = {Phys. Rev. B},
  year      = {1999},
  volume    = {59},
  pages     = {3969-3980},
}

@ARTICLE{dijkstra59,
  author = {Dijkstra, E. W.}, 
  journal = {Numerische Math.}, 
  volume = {1}, 
  pages = {269-271}, 
  year = {1959}, 
  title = {A note on two problems in connexion with graphs} 
}

@article{kumedamw01,
  author   = {Kumeda, Y. and Munro, L. J. and Wales, D. J.},
  title    = {Transition States and Rearrangement Mechanisms from Hybrid Eigenvector-Following and
              Density Functional Theory. Application to {C$_{10}$H$_{10}$} and Defect Migration in Crystalline Silicon},
  journal  = {Chem. Phys. Lett.},
  volume   = {341},
  year     = {2001},
  pages    = {185-194}
}

@ARTICLE{CarrTW05,
  author    = {Carr, J. M. and Trygubenko, S. A. and Wales, D. J.},
  title     = {Finding pathways between distant local minima},
  journal   = {J. Chem. Phys.},
  year      = {2005},
  volume    = {122},
  pages     = {234903}
}

@article{Wales04,
  title={Some further applications of discrete path sampling to cluster isomerization},
  author={Wales, David J},
  journal={Molecular physics},
  volume={102},
  number={9-10},
  pages={891--908},
  year={2004},
  publisher={Taylor \& Francis}
}

@article{Wales18,
author = {Wales, David J.},
title = {Exploring Energy Landscapes},
journal = {Annu. Rev. Phys. Chem.},
volume = {69},
pages = {401-425},
year = {2018},
}

@article{Wales02,
  title={Discrete path sampling},
  author={Wales, David J},
  journal={Molecular physics},
  volume={100},
  number={20},
  pages={3285--3305},
  year={2002},
  publisher={Taylor \& Francis}
}

@article{FrodasonVJV23,
  title={Migration of {Ga} vacancies and interstitials in {$\beta$-Ga$_2$O$_3$}},
  author={Frodason, Ymir K and Varley, Joel B and Johansen, Klaus Magnus H and Vines, Lasse and Van de Walle, Chris G},
  journal={Physical Review B},
  volume={107},
  number={2},
  pages={024109},
  year={2023},
  publisher={APS}
}

\end{document}